\documentclass[dvips,12pt]{article}
\usepackage{graphicx}
\begin{document} 
\thispagestyle{empty} 

\begin{center}
{\large{\bf APPLICATION OF THE GIBBS EQUILIBRIUM CONDITIONS\\
TO THE QGP-HADRON TRANSITION CURVE\\}}
\vspace{2cm} 
{\large A.~S.~Kapoyannis}\\ 
\smallskip 
{\it Department of Physics, University of Athens, 15771 Athens, Greece}\\
\vspace{1cm}

\end{center}
\vspace{0.5cm}
\begin{abstract}
A method is developed to consistently satisfy the Gibbs equilibrium
conditions between the quark-gluon and hadronic phase, although each
phase has been formulated in separate grand canonical partition
functions containing three quark flavours. The sector in the space of
thermodynamic variables, where the transition takes place, is restricted
to a curve, according to the phase diagram of QCD. The conservation laws
of quantum numbers are also imposed on the transition curve. The effect of
the inclusion of the newly discovered pentaquark states is considered.
The freeze-out conditions of $S+S$, $S+Ag$ (SPS) and $Au+Au$ (RHIC) are
found compatible with a primordial QGP phase, but the conditions indicated by $Pb+Pb$ (SPS) are not.
\end{abstract}

\vspace{3cm}

PACS numbers: 12.40.Ee, 05.70.Fh, 12.38.Mh, 24.10.Pa

Keywords: critical line, QGP-hadron transition,
partial chemical equilibrium fugacities, Gibbs equilibrium, heavy-ion

{\it e-mail address:} akapog@phys.uoa.gr (A.S. Kapoyannis)

\newpage
\setcounter{page}{1}

\vspace{0.3cm}
{\large \bf 1. Introduction}

Quantum Chromodynamics is universally accepted as the theory
of strong interactions. Within the context of this
theory the phase of Quark-Gluon plasma receives accurate description.
However, the formation of the hadronic phase, which is the final state of
any possible primordial QGP state, still remains an open
problem in view of QCD. On the other hand, the hadronic multiplicities
emerging from heavy-ion collisions have been extensively and successfully
predicted by statistical models using a handful of thermodynamical
parameters [1-8]. So the use of two separate models for the QGP and the
hadron phase, called Hadron Gas (HG), offers a complementary approach.

QCD predicts that the transition between QGP and the hadronic phase is a
first order one at high baryon densities (depicted by a critical line
on the ($T,\mu_B$) plane), while it is of a higher order at small or zero
baryon densities (crossover). The end point of the first order transition
line is a critical point [9]. The transition  points must be restricted
to a curve on the phase diagram of temperature and baryon chemical
potential. In view of this aspect any models to be used for the
description of QGP and HG have to be matched properly at the transition
between the two phases.

The aim of this work is to trace the sector of the space of thermodynamic
variables, where the QGP-hadron transition occurs, with the following
requirements:
a) Any mixed phase formed in the first order part of the transition must
occupy only a curve in the space of the thermodynamic variables. This
requirement is even stronger in the crossover area where a mixed phase
does not exist.
b) The Gibbs equilibrium conditions have to be satisfied, which amount to
$T_{QGP}=T_{HG}$ for thermal equilibrium, $P_{QGP}=P_{HG}$ for mechanical
equilibrium and
$\{\mu\}_{QGP}=\{\mu\}_{HG}$ for chemical equilibrium, where $\{\mu\}$
stands for the set of chemical potentials used in the description of the
two phases.
c) All the conservation laws of quantum numbers like baryon number $B$,
electric charge $Q$, strangeness $S$, etc. have to be satisfied at every
point on the transition line, in a way that they could be extended for
every number of flavours that are present in the system.

These problems are confronted every time separate partition functions are
used for the two phases, but the simultaneous fulfilment of the above
conditions has not been achieved.
Among the numerous examples that exist, in [10], where only light,
identical quarks are used ($u=d \equiv q$), the curves of equal pressures
are made to approximately coincide by a choice on the external parameters
$B$ and $a_s$, something which does not allow matching when other
flavours are introduced. In [11] the strange fugacity $\lambda_s$ is
discontinuous at the HG-QGP transition and the conservation of baryon
number can only be accommodated in the case of first order transition.
In [12] the strange chemical potential $\mu_s$ is also discontinuous. In
[13] only $q$ quarks are considered and the requirement of continuity of
chemical potentials and conservation of the baryon number leads to a mixed
phase which occupies a surface and not a line on the ($T,\mu_B$) plane.
The same is true in [14-16] where also $s$ quarks are included.
In [17] there is an analogous situation as in [13] with a critical line
at the ($T,\mu_B$) plane but the conservation of baryon number is not
considered.
In [18] the $q$ and $s$ quark chemical potentials are continuous but
baryon number and strangeness of the system are not kept constant during
hadronisation, since hadrons evaporate from QGP. The considerations of [11-18] are consistent with a first order transition but cannot be valid
at the crossover region.
In this work all the thermodynamic variables and the pressure will be kept
continuous at the transition line (in contrast with [10-12]), the first
order part of the transition will be presented by a line on the
($T,\mu_B$) plane (differing from [13-16]), in the mixed phase the
quantum numbers will be conserved to each constituent phase (differing
from [14]) and no evaporation of hadrons will be assumed from the system
(differing from [18]).

Let us consider the requirements that a system with $N_f$ quark
flavours has to satisfy. Every
conservation law accounts for two equations to be fulfilled. One
sets the value of the quantum quantity, e.g. $\left<B\right>_{QGP}=b_i$ and the other
assures the conservation at the phase transition, e.g. $\left<B\right>_{QGP}=\left<B\right>_{HG}$.
The total number of equations that must hold are, thus, $2N_f+1$ (the unit
accounts for the equality of pressures). Assuming the existence of full chemical equilibrium,
every quark flavour introduces one extra fugacity in the set of the
thermodynamical variables, which, with the inclusion of volume and
temperature, amount to $N_f+2$. At the crossover region, the surviving
free parameters required to fulfil the necessary equations decrease to $N_f+1$,
because of the equality of densities and consequently the equality of volumes between
the two phases ($V_{QGP}=V_{HG}$). At the first order
transition line the free parameters are $N_f+2$, since now
$V_{QGP} \neq V_{HG}$. It is evident then that the necessary $2N_f+1$
conditions can be fulfilled only at the first-order part of the transition
and only when there is one flavour present, $N_f=1$, or when the $u$ and $d$
quarks are considered identical ($q$ quarks, described by a single chemical
potential $\mu_q$). It has to be clarified that the conditions like
$\left<B\right>_{QGP}=b_i$ have to be satisfied in order to have a whole line of
transition points. If these equations are dropped, then we are left with
$N_f+1$ equations, which can be solved, but result to a unique point in
the space of the thermodynamical parameters.

\vspace{0.3cm}
{\large \bf 2. Expanding the fugacity sector}

It is clear that in order to satisfy $2N_f+1$ relations, every
flavour has to be accompanied by two fugacities instead of one.
The multiplicity data emerging from heavy ion collisions suggest that the
thermalised hadronic system has not achieved full chemical equilibrium.
First the strangeness partial chemical equilibrium factor $\gamma_s$ had
been introduced [2] and used extensively to model the data [3-4]. Also a
similar factor for the light quarks $\gamma_q$ was introduced [5] and
used in certain analyses [6]. Here the light $u$ and $d$ quarks will be
accompanied by separate fugacities $\gamma_u$, $\gamma_d$. A factor
$\gamma_j$ controls the quark density \mbox{$n_j\!+\!n_{\bar{j}}$} in
contrast with the usual fugacity $\lambda_j$ which controls the net quark
density $n_j\!-\!n_{\bar{j}}$ [3]. These additional fugacities can serve
the purpose of satisfying the necessary equations at the transition point, as well as, preserving the continuity of chemical potentials
between the two phases.

A system with 3 flavours ($u$, $d$ and $s$ quarks)
is described by the set of thermodynamical variables
$(T,\lambda_u,\gamma_u,\lambda_d,\gamma_d,\lambda_s,\gamma_s)\equiv$
$(T,\{\lambda,\gamma\})$.
Setting $x=V_{HG}/V_{QGP}$, the set of
equations to be satisfied at every phase transition point will be
\begin{equation}
P_{_{QGP}}(T,\{\lambda,\gamma\})=P_{_{HG}}(T,\{\lambda,\gamma\})
\end{equation}
\begin{equation}
n_{B_{QGP}}(T,\{\lambda,\gamma\})=x\;n_{B_{HG}}(T,\{\lambda,\gamma\})
\end{equation}
\begin{equation}
n_{Q_{QGP}}(T,\{\lambda,\gamma\})=x\;n_{Q_{HG}}(T,\{\lambda,\gamma\})
\end{equation}
\begin{equation}
n_{S_{QGP}}(T,\{\lambda,\gamma\})=x\;n_{S_{HG}}(T,\{\lambda,\gamma\})
\end{equation}
\begin{equation}
n_{B_{QGP}}(T,\{\lambda,\gamma\})=2\beta\;n_{Q_{QGP}}(T,\{\lambda,\gamma\})
\end{equation}
\begin{equation}
n_{S_{QGP}}(T,\{\lambda,\gamma\})=0\;,
\end{equation}
where $n$ denotes densities.
For isospin symmetric systems one has to set $\beta=1$ in (5).
Eqs.~(1)-(6) only have one free variable, necessary to produce a whole transition line in the phase diagram. At crossover $x=1$, whereas at the
first order transition line the inequality $V_{QGP} \neq V_{HG}$
preserves the survival of $x$ as an extra variable.

\vspace{0.3cm}
{\large \bf 3. A solution for the transition curve}

The above considerations are applicable to every
partition function connected to the hadronic and the quark state.
It is interesting, though, to check whether they can produce a real solution
for the transition curve, i.e.~that the system of equations (1)-(6) is not
impossible. For this reason two simple models, will be employed.
For the Hadron Gas phase only the repulsive part of the hadronic interaction
will be taken into account through a Van der Waals treatment of the system
volume. Also the correct Bose-Einstein or Fermi-Dirac statistics
applicable to each hadron will be considered.
The hadronic partition function will be extended to all hadronic states
containing $u$, $d$ and $s$ quarks as they are listed in [19].
First, the HG partition function for point particles can be written down as
\begin{equation}
\ln Z_{HG\;pt}(V,T,\{\lambda,\gamma\})=
\frac{V}{6\pi^2 T}
\sum_{\rm a} \sum_i g_{{\rm a}i}
\int_0^{\infty} \frac{p^4}{\sqrt{p^2+m_{{\rm a}i}^2}}
\frac{1}{e^{\sqrt{p^2+m_{{\rm a}i}^2}/T}\lambda_{\rm a}^{-1}+\alpha}dp\;,
\end{equation}
where $g_{{\rm a}i}$ are degeneracy factors due to spin and isospin and
$\alpha=-1(1)$ for bosons (fermions).
Index ${\rm a}$ runs over all hadronic families, each of which contains
members with the same quark content and $i$ over all the particles of this
family. The fugacity
$\lambda_{\rm a}=\prod_j\lambda_j^{N_j-N_{\bar{j}}} \gamma_j^{N_j+N_{\bar{j}}}$,
where $j=u,d,s$ and $N_j(N_{\bar{j}})$ is the number of $j(\bar{j})$ quarks
contained in a hadron belonging to family ${\rm a}$. 
For the light unflavoured mesons with quark content
$(c_1/2)(u\bar{u}+d\bar{d})+c_2 s\bar{s}$, $c_1+c_2=1$, the fugacity used is
$\lambda_{\rm a}=(\gamma_u\gamma_d)^{c_1}\gamma_s^{2c_2}$.

If each hadron $i$ is assumed to occupy volume $V_i$, then the available
volume for the system is reduced to $\Delta=V-\sum_i N_i V_i$, where $N_i$ is
the number of particles $i$ present in the system.
Assuming that each hadrons' volume is proportional to its mass, then
$V_{i}/m_{i}=V_0$,
with $V_0$ remaining an open parameter controlling the hadron size.
The available volume can be written as
$\Delta=V-\sum_i N_i m_i V_0$. Dividing by $\Delta$ we get
\[
1=\frac{V}{\Delta}-\rho_{pt} V_0 \Rightarrow
\Delta=\frac{V}{1+\rho_{pt} V_0}\;,
\]
where $\rho_{pt}$ is the mass density of a system of point particles at
volume $\Delta$. Defining the quantity $f\equiv\rho_{pt} V_0$, it can be
calculated to be
\begin{equation}
f=\frac{V_0}{6\pi^2 T}
\sum_{\rm a} \sum_i g_{{\rm a}i}m_{{\rm a}i}
\int_0^{\infty} \frac{p^4}{\sqrt{p^2+m_{{\rm a}i}^2}}
\frac{1}{e^{\sqrt{p^2+m_{{\rm a}i}^2}/T}\lambda_{\rm a}^{-1}+\alpha}dp\;,
\end{equation}
where the sum over all particles is organised first to a sum over all
families. The available volume is then
\begin{equation}
\Delta=\frac{V}{1+f}\;.
\end{equation}

The partition of the
extended hadrons at volume $V$ may be considered equal to the partition
function of the point particles at volume $\Delta$
\begin{equation}
\ln Z(V,T,\{\lambda,\gamma\})=\ln Z_{pt}(\Delta,T,\{\lambda,\gamma\})\;.
\end{equation}
The pressure is 
\[
P_{HG}=\frac{\partial \ln Z_{HG}(V,\ldots)}{\partial V} =
\frac{\partial \ln Z_{HG\;pt}(\Delta,\ldots)}{\partial V}\;,
\]
or by using eq.~(9)
\begin{equation}
P_{HG}=\frac{P_{HG\;pt}}{1+f}\;.
\end{equation}
In a similar manner the densities are calculated to be
\begin{equation}
n_{i\;HG}=\frac{n_{i\;HG\;pt}}{1+f}
\end{equation}

For the QGP phase a simple model containing 3 flavours is used. The quarks
are non-interacting and only the presence of gluons is accounted for, as
well as the effect of the vacuum through the MIT bag constant, $B$. A
wealth of quark fugacities is easily accommodated, though, in this model.
The QGP partition function is consequently
\begin{equation}
\ln Z_{QGP}(V,T,\{\lambda,\gamma\})=
\frac{N_s N_c V}{6\pi^2 T}
\sum_j \int_0^{\infty} \frac{p^4}{\sqrt{p^2+m_j^2}}
\frac{1}{e^{\sqrt{p^2+m_j^2}/T}(\lambda_j \gamma_j)^{-1}+1} dp
+V\frac{8\pi^2 T^3}{45}-\frac{BV}{T}\;,
\end{equation}
where $N_s=2$ and $N_c=3$.
Index $j$ runs to all quarks and antiquarks and the fugacity 
$\lambda_{\bar{j}}=\lambda_j^{-1}$ and $\gamma_{\bar{j}}=\gamma_j$. The
current quark masses are $m_u=1.5$, $m_d=6.75$ and $m_s=117.5$ MeV [19].

At the first order QGP-HG transition a mixed phase is assumed. This phase
spans over a curve in the ($T,\{\lambda,\gamma\}$)-space, so these
variables are kept constant throughout the mixed phase. The only
thermodynamic variable which is allowed to change is the system volume $V$.
The volume equals $V_{HG}$ at the pure hadronic phase, at one end of the
first order transition and $V_{QGP}$ at the pure quark phase, at the other
end of the transition. The partition function of the mixed phase can then
be written down as
\begin{equation}
\ln Z_{mixed}(V,T,\{\lambda,\gamma\})=
\delta \ln Z_{HG}(V_{HG},T,\{\lambda,\gamma\})+
(1-\delta) \ln Z_{QGP}(V_{QGP},T,\{\lambda,\gamma\})\;.
\end{equation}
The parameter $\delta$ is 
$0 \le \delta \le 1$ and for $\delta=1(\delta=0)$ we have pure HG(QGP)
phase. Since in all cases the volume appears as a multiplicative factor
in the partition function, the ratio $g\equiv\frac{\ln Z}{V}$ does not
depend on the volume. Then eq.~(1) suggests that at every transition point
\begin{equation}
g_{HG}(T,\{\lambda,\gamma\})=g_{QGP}(T,\{\lambda,\gamma\})\;.
\end{equation}
Then eq.~(14) is equivalent to
\begin{equation}
\ln Z_{mixed}(V,T,\{\lambda,\gamma\})=
\left[\delta V_{HG}+(1-\delta) V_{QGP}\right] g_{HG}(T,\{\lambda,\gamma\})\;,
\end{equation}
suggesting that the system volume $V$ of the mixed phase can be defined as
\begin{equation}
V=\delta V_{HG}+(1-\delta) V_{QGP}\;.
\end{equation}
Through eqs.~(1), (16) and (17), the pressure in the mixed phase can be
calculated for every $V$ to be
\begin{equation}
P_{mixed}=\frac{kT \ln Z_{mixed}(V,T,\{\lambda,\gamma\})}{V}=
kTg_{HG}(T,\{\lambda,\gamma\})=P_{HG}=P_{QGP}\;.
\end{equation}
The pressure is, consequently, kept constant throughout the first order
transition. The part of the $P\!-\!V$ isotherm which corresponds to the mixed phase is parallel to the $V$ axis, as it is expected in a first
order transition.

On the contrary, the densities vary. Let $\lambda_i$ be the fugacity
associated with the density $n_i$. Then
\[
n_{i\;mixed}=\frac{1}{V}\lambda_i
\frac{\partial \ln Z_{mixed}(V,T,\{\lambda,\gamma\})}{\partial \lambda_i}=
\hspace{5cm}
\]
\[
\hspace{2cm}
\frac{1}{V}\left[
\delta V_{HG} \lambda_i \frac{\partial g_{HG}(T,\{\lambda,\gamma\})}{\partial \lambda_i}+
(1-\delta) V_{QGP} \lambda_i \frac{\partial g_{QGP}(T,\{\lambda,\gamma\})}{\partial \lambda_i}
\right]\;,
\]
or
\begin{equation}
n_{i\;mixed}=\frac{\delta V_{HG}}{\delta V_{HG}+(1-\delta) V_{QGP}}n_{i\;HG}+
\frac{(1-\delta) V_{QGP}}{\delta V_{HG}+(1-\delta) V_{QGP}}n_{i\;QGP}\;.
\end{equation}
It is easily checked that for $\delta=1$($\delta=0$) the density of the
pure hadronic(quark) state is produced.
The conservation of the quantum quantities $B$, $Q$ and $S$ is assured in
the mixed phase. Eq.~(2) implies that
$\left<B\right>_{HG}=\left<B\right>_{QGP}$.
Then multiplying the two sides of eq.~(19) by $V$ gives
\begin{equation}
Vn_{B\;mixed}=
\delta\left<B\right>_{HG}+(1-\delta)\left<B\right>_{QGP}
\Rightarrow
\left<B\right>_{mixed}=\left<B\right>_{HG}=\left<B\right>_{QGP}\;.
\end{equation}
Eqs.~(3)-(4) produce similar equations for $Q$ and $S$.

Between the crossover region (where $x=1$) and the first order transition
line (where $x\neq1$) the critical point resides. Observing eqs.~(1)-(6)
it is clear that they do not provide a restriction on $x$, other than it
should be a continuous function. So these equations can accommodate
an additional constraint in the form of $x$, which may be provided by
a sophisticated partition function that records the full part of
interaction (attractive part as well). 

The system of eqs.~(1)-(6) can then be solved for $x=1$ for the crossover
region or for $x\neq1$ for the first order transition curve.
The system is simplified observing that the strangeness neutrality at the
QGP phase (eq.~(6)) leads to the solution $\lambda_s=1$. This solution is
valid for every case of QGP partition function, as long as products of
the fugacities of the strange quark with the fugacities of $u$ or $d$
quarks do not appear. The initial system then is reduced to the system of
eqs.~(1)-(5) for $\lambda_s=1$. The HG partition function (7), (10) and
the QGP partition function (13) is used to the system of eqs.~(1)-(5).
For the particular choices of partition functions, two parameters, $B$ and
$V_0$, are left open, producing different solutions for the transition
curves. The system of eqs.~(1)-(5) for $\lambda_s=1$ accepts as solution
for the variable $\gamma_s$ the value 0, since then eq.~(5) is
automatically satisfied.
This a trivial solution because it is equivalent to the absence of the
strange quarks in the system and therefore such solutions should be excluded. Non-trivial solutions for the thermodynamic variables are
depicted for the parameters $B^{1/4}=280$ MeV and $V_0=1.4/(4B)$ in Figs.~1-4 for the isospin symmetric case ($\beta=1$). Lines (a) represent the solution without the inclusion of pentaquarks. The crossover region,
which is determined uniquely after setting the parameters $B$ and $V_0$,
is drawn everywhere with slashed lines. 

Two additional matters concerning the phase transition
are the position of the critical point where the crossover region ends and
the first order transition sets in, as well as, the ratio of the volumes
$x=V_{HG}/V_{QGP}$ at the first order transition line. These two matters
cannot be dealt with the simple choices of partition functions used for
the calculations of this section, since the attractive part of the
interaction among hadrons or quarks is completely neglected.
To display certain solutions for the critical curve within the context of
the partition functions (7), (10) and (13), a position for the critical
point has to be chosen. This position is set at $\mu_{B\;cr.p.}=360$ MeV, according to the result of [20], where the critical point is located by lattice QCD calculations using three quark flavours and considerably reduced light quark masses, approaching their physical values.
Eqs.~(1)-(6) for $x=1$ is a system of 6 equations with 7 variables. 
Adding one more equation ($\mu_{B\;cr.p.}=360$ MeV) a system with 7 equations is formed, which is solved to determine the full set of thermodynamic variables that correspond to the position of the critical point.

A form for the ratio of the volumes $x$ also has to be defined. This form has to produce $x=1$ at the position of the critical point. Moreover, it is chosen to produce a given value $x_1$ at a specific value of
$\lambda_{u\;1}$. A simple form which implements these demands is
\begin{equation}
x=1+\left(\frac{\ln\lambda_u-\ln\lambda_{u\;cr.p.}}
{\ln\lambda_{u\;1}-\ln\lambda_{u\;cr.p.}}\right)^{\epsilon}(x_1-1)\;,
\end{equation}
where the exponent $\epsilon$ regulates the curvature of
the first order transition line. For lines (a) $x_1=1.1$ 
at $\lambda_{u\;1}=14.2$ and $\epsilon=1.5$ are chosen. Of course, any
function of $x$ can be used, producing different first order transition
curves.
The resulting first order transition lines are drawn with solid lines in
Figs~1-4, while the respective critical points are represented by solid circles. 

Temperature $T$ is displayed as function of the baryon chemical potential $\mu_B$ in Fig.~1. In the same figure, line (d), which represents the phase transition line as it is calculated from the Lattice QCD in [21], is drawn for comparison. 
The relative chemical equilibrium fugacity $\gamma_u$ is displayed as
function of $\mu_B$ in Fig.~2. This particular
solution leads to the gradual suppression of $\gamma_u$ as baryon chemical
potential increases. The connection of $\gamma_u$ and $\gamma_d$, for
isospin symmetric solution, is depicted in Fig.~3. The line
\mbox{$\gamma_u=\gamma_d$} is also drawn for comparison.
The relative chemical equilibrium factor $\gamma_s$ is drawn as a
function of the baryon chemical potential in Fig.~4.

In Fig.~5 the ratios of volumes $x$, which are used in the first order
transition, are drawn against the baryon chemical potential.
The used forms of $x$ are increasing functions with respect to the baryon chemical potential. The resulting first order transition lines produce
smaller temperatures as the baryon chemical potential increases, something
which is expected.

One direct consequence of the simultaneous solution of eqs.~(1)-(6) is
that the relative chemical equilibrium fugacities have values that depend
on each other at every transition point. This is easily realised by
inspecting the condition $n_{S_{HG}}=0$ (for $\lambda_s=1$). The solution
of this condition is greatly simplified by the use of the Boltzmann
approximation and the assumption that
isospin symmetry leads to the approximate solution
$\lambda_u=\lambda_d\equiv\lambda_q$ and $\gamma_u=\gamma_d\equiv\gamma_q$.
Neglecting trivial solutions, where $\gamma_s = 0$, the zero strangeness
condition can be solved to give
\begin{equation}
\gamma_s=
\frac{F_K(T)-F_H(T)\gamma_q(\lambda_q+\lambda_q^{-1})}{2F_{\Xi}(T)}\;.
\end{equation}

In eq.~(22), $F_K$ corresponds to the Kaon mesons, $F_H$ to the Hyperon
baryons ($\Lambda$'s and $\Sigma$'s) and $F_{\Xi}$ to the $\Xi$ baryons,
while the summation
\begin{equation}
F_{\rm a}(T)=\frac{T}{2\pi^2}
\sum_i g_{{\rm a}i} m_{{\rm a}i}^2 K_2(\frac{m_{{\rm a}i}}{T})
\end{equation}
to all particles $i$ of the same family is implied.
In the above relation, $K$ denotes a modified Bessel function of the second
kind. It is evident from eq.~(22) that the increase of the relative
chemical equilibrium factor for light quarks, $\gamma_q$ and the increase
of the light quark fugacity, $\lambda_q$, leads, at constant temperature, to the decrease of factor $\gamma_s$.

\newpage
\vspace{0.3cm}
{\large \bf 4. Inclusion of pentaquarks}

There has been recent evidence of hadrons containing five quarks. These
5-quark states are the $\Theta^+(1540)$ [22] with
$I=0$ and quark content $uudd\bar{s}$ and the $\Xi^*(1862)$ with $I=3/2$.
The content of the states $\Xi^*(1862)$ is $ssdd\bar{u}$ (for the state with
electric charge Q=-2), $ssud\bar{u}$ (with Q=-1), $ssud\bar{d}$ (with Q=0)
and $ssuu\bar{d}$ (with Q=+1). The existence of the first three of the states
$\Xi^*(1862)$ has been confirmed [23]. 
The motivation to investigate the effect of the 
pentaquark states comes from the fact that these states can alter eqs.~(1)-(6), since they introduce additional hadronic families, each of which is accompanied by completely different functions between the system fugacities compared to the ones in the already known families. This can
be easily realised if the corresponding equation of (22) is written down
as
\begin{equation}
\gamma_s=
\frac{F_K(T)+F_{\Theta}(T)\gamma_q^3(\lambda_q^2+\lambda_q^{-2})
(\lambda_q+\lambda_q^{-1})
-F_H(T)\gamma_q(\lambda_q+\lambda_q^{-1})}
{2[F_{\Xi}(T)+F_{\Xi^*}(T)\gamma_q^2]}\;.
\end{equation}
The existence of $\Theta$ hadron drives $\gamma_s$ to higher values with a
strong dependence on $\gamma_q$ and $\lambda_q$, whereas the inclusion of
the $\Xi^*$ states contributes to the decrease of $\gamma_s$.

The system of eqs.~(1)-(6) is then solved with the inclusion of the
$\Theta^+(1540)$ and $\Xi^*(1862)$ states with the same partition functions for the HG and the QGP phase and for the same parameters $B$, $V_0$ as in the case without the inclusion of the pentaquarks (lines (a) of the previous section). The resulting curves are lines (b), shown in Figs.~1-4.
The adopted form of the ratio of volumes $x$ again produces value
$x_1=1.1$ at $\lambda_{u\;1}=14.2$ (while $\epsilon=1.5$) and is plotted in Fig.~5.

Lines (a) and (b) record the difference induced in the transition curve
by the inclusion of the pentaquarks, if parameters $B$ and $V_0$ remain
the same and $x$ acquires the same value at a given value of $\lambda_u$.
However, none of these parameters is known and so the difference in the transition curve cannot provide evidence for the existence of pentaquarks.
For this reason lines (c) are drawn in Figs.~1-4. These lines represent a
solution for the transition curve without the inclusion of the pentaquarks
but for a different choice of parameters ($B$ remains the same, $V_0=1.29/(4B)$, $x_1=1.12$ at $\lambda_{u\;1}=14.2$ and $\epsilon=1.15$).
It is evident, now, that lines (b) (with pentaquarks included) and
lines (c) (without pentaquarks) almost coincide.

\vspace{0.3cm}
{\large \bf 5. Application to heavy-ion data}

The fact that the transition territory between the hadronic and the
QGP phase is restricted to a line in the space of temperature and chemical
potentials produces a direct connection between the variables after the
phase transition and the corresponding ones before: they {\it must coincide}.  
This is not the case when the transition territory is allowed to occupy a surface. Then the connection between the thermodynamic variables of the hadronic and the quark phase is broken. The system, as it
crosses the transition territory to enter the state of hadrons, loses its ``memory'' of the plasma state.

In a system where the hadronic state carries the memory of its preceding state, one may use the freeze-out variables as a diagnostic
tool for a primordial QGP phase. Assuming that (a) a quark-gluon phase {\it has}
been formed in a collision experiment and (b) the chemical freeze-out
occurs right after the transition to the hadronic phase, then the freeze-out
thermodynamic variables have to fulfil constraints (1)-(6).
If, on the contrary, no quark-gluon state is formed before hadronization,
then the restriction on the freeze-out conditions of the system is
diminished only to eqs.
\begin{equation}
n_{B_{HG}}(T,\{\lambda,\gamma\})=2\beta\;n_{Q_{HG}}(T,\{\lambda,\gamma\})
\end{equation}
\begin{equation}
n_{S_{HG}}(T,\{\lambda,\gamma\})=0\;,
\end{equation}
a set of constraints which will be referred to from now on as set A.

The thermodynamic variables are
extracted through a fit of the experimentally measured particle
multiplicities or ratios to a statistical model. Such a technique has
been successful. If now the additional constraints (1)-(6) are
imposed, the question that arises is whether a successful fit is also
produced or the restrictions that these constraints imply are inconsistent with the data.

These ideas will now be used to analyse the freeze-out variables of four
heavy-ion experiments. The application of eqs.~(1)-(6) require the
knowledge of the partition functions for the hadron and the quark phase. The particular functions used in sections 4 and 5 to demonstrate a solution include the arbitrariness in the choice of the parameters $B$,
$V_0$ and quantity $x$.
Since it is unwanted for the extracted variables to depend on the choice
of unknown parameters, it is better to form and apply a subset of constraints which are completely independent from these parameters.

When the system of eqs.~(1)-(6) is valid, eqs.~(3)-(6) can equivalently be rewritten as
\begin{equation}
n_{B_{HG}}(T,\{\lambda,\gamma\})=2\beta\;n_{Q_{HG}}(T,\{\lambda,\gamma\})
\end{equation}
\begin{equation}
n_{B_{QGP}}(T,\{\lambda,\gamma\})=2\beta\;n_{Q_{QGP}}(T,\{\lambda,\gamma\})
\end{equation}
\begin{equation}
n_{S_{HG}}(T,\{\lambda,\gamma\})=0\;,
\end{equation}
\begin{equation}
n_{S_{QGP}}(T,\{\lambda,\gamma\})=0\;.
\end{equation}
Eq.~(27) results from eqs.~(2), (3) and (5), whereas eq.~(30) is due to  eqs.~(4) and (6). It is easily checked that the common denominator (see eq.~(12)), which includes the parameter $V_0$, cancels out from both sides of eq.~(27). The same is true for eq.~(29), while eq.~(30) is
easily solved to give $\lambda_s=1$. Eqs.~(27)-(30) (referred to as Set
B) now form a set of equations that do not depend on
the parameters $V_0$ for the particle size, $B$ (MIT bag constant) nor the
ratio $x$ applicable to the first order transition line.

On the contrary, eqs.~(1) and (2) are model dependent and contain unknown
parameters. However, if the freeze-out parameters are determined, they can
be inserted in eq.~(2) to determine $V_0$ (assuming that $x$ is known) and
then eq.~(1) can be used to determine $B$. This task serves to show that
eqs.~(1) and (2) have a real solution and contributes to the overall
consistency of the technique.

In the following the freeze out variables for the experiments $S+S$ [24],
$S+Ag$ [25] (NA35) at beam energy $200$ AGeV, $Pb+Pb$ [26] (NA49) at beam
energy $158$ AGeV and $Au+Au$ [27] (STAR) at $\sqrt{s_{NN}}=130$ GeV will
be analysed. The data used are listed in Table 1 and they are in all cases
 full phase space multiplicities, except from
the RHIC data which are measured in the midrapidity. The experiments to be analysed are chosen because they do not produce great baryon chemical potential at freeze out and so they are probably at the crossover area [20], allowing one to set $x=1$. The technique can be applied to the first order transition case, determining the freeze-out variables, since the equations of set B do not depend on $x$, but then the parameters
$V_0$ and $B$ cannot not be uniquely determined.

The theoretical calculation of the particle multiplicity necessary to
perform a fit to the experimental data has been carried out with the
partition function (7), (10). The right
Bose-Einstein or Fermi-Dirac statistics for every particle has been used
throughout the calculations. The feeds from the decay of resonances have also been included.
The value of $\beta$
is set to 1 in the case of $S+S$, 1.1 in the case of $S+Ag$, 1.27 for
$Pb+Pb$ and 1.25 for $Au+Au$.

In Tables 2 (a)-(b) the freeze-out variables are extracted with the
assumption of no primordial QGP phase (constraints of set A), whereas
in Tables 3 (a)-(b) the primordial QGP phase is assumed, thus applying
constraints of set B. Another matter concerning the analysis is the observation that the inclusion of the pion multiplicity deteriorates the fit [8,28]\footnote{The
presence of excess of pions, though, can be connected with a primordial
high entropy phase or with a phase with the chiral symmetry restored [29].}. 
Since the quality of the fitting procedure is of importance in evaluating the results and a bad fit, when the constraints of set B are imposed, may be partly due to the presence of the pion
multiplicity, two fits are performed in each case, one with all the multiplicies included (Tables (a)) and one without the multiplicities that contain pions (Tables (b)). This makes clear the effect of the pion multiplicities in the overall procedure. In any case, the fits without the pions are in general more reliable.

The first observation which can be made by comparing the first and the
second part of every table is that the inclusion of pentaquarks has negligible effect on the evaluated parameters or the quality of the fit.
So one can safely draw equivalent conclusions by performing the analysis with the pentaquarks or without them.

For the $S+S$ and $S+Ag$ data the fit with set B is of medium
quality ($\chi^2/dof=2.95$ and $1.92$, respectively) when the pions are
present. This is not far worse, though, than the fit in the case of set A. When
the pions are excluded, the fit with set B turns out to be very good 
($\chi^2/dof=0.47$ and $0.065$, respectively), while the temperature remains at
acceptable values, proving these cases to be completely compatible with a primordial
quark-gluon phase.
Another observation is that the imposition of set B with respect to set A, in every case, does not produce a dramatic change in the quality of the fit.

In the case of $Pb+Pb$ the fit is relatively good with the imposition of set A ($\chi^2/dof=2.50-1.79$), but the imposition of set B severely worsens the quality of the fit. The situation cannot be remedied with the exclusion of pions and $\chi^2/dof$ remains at the value of $18.0-18.2$. 
The conclusion to be drawn from the bad quality of the fit with the imposition of the constraints of set B, is that the data of this experiment are not compatible with a preceding QGP phase.
The dramatic change in the quality of the fit between the cases of set A and B, should be noted as well.
Also the freeze-out temperature in case of set B is unrealistically high and rises enormously with respect to the case of set A.

The findings concerning the $S+S$ and $S+Ag$ data are also in agreement 
with the proximity of the chemical freeze out points of these experiments to the Statistical Bootstrap critical line which was found in [8]. On the
contrary, the freeze out point of $Pb+Pb$ was not found to possess such an
attribute in [30], which is also in agreement with the present results.

In case of RHIC and in case of set A the value of $\chi^2/dof$ is $1.51$,
when the pions are included and this value diminishes to $0.229$, when the
pions are excluded. Similar results are obtained when set B is imposed.
The fit in the presence of pions is of not so good quality ($\chi^2/dof=3.86$) and the
temperature aquires too high value. The fit turns out to be quite good, though, when
the pions are excluded ($\chi^2/dof=1.19$) and the temperature remains at acceptable
values, so the thermodynamic parameters are compatible with a QGP phase.

The extracted parameters in case of set B are inserted to eqs.~(1) and (2)
and the parameters $V_0$ and $B^{1/4}$ are also determined. It is interesting
that in the cases of $S+S$, $S+Ag$ and $Au+Au$ (without the pions), which
have been proven to be compatible with set B, all the calculated values of $V_0$
and $B^{1/4}$ are close, compatible with a unique value for these parameters. On the contrary, the $Pb+Pb$ case produces values of $V_0$
and $B^{1/4}$ with no connection with the rest cases.

The necessity of the expansion of the fugacity sector with the partial
equilibrium fugacities is also revealed with the application of the
present technique. If these fugacities are set to
$\gamma_u=\gamma_d=\gamma_s=1$, then the sector of the phase space which
is compatible with the QGP-hadron transition is severely limited.
In such a case, if a similar fit to the one with set B is performed, apart from the
fact that eqs.~(27) and (28) cannot be accommodated, the fit turns out to
be worse. The result in the case without the pions is then $\chi^2/dof=0.617$, $1.09$,
$28.0$ and $1.83$ for $S+S$, $S+Ag$, $Pb+Pb$ and $Au+Au$ respectively.
Then in the case of $Au+Au$ the compatibility with the QGP phase turns out to be dubious.

Another general observation, which can be drawn from Tables 2-3, is that in the cases
which are compatible with a primordial QGP phase ($S+S$, $S+Ag$ and $Au+Au$) the
inclusion of the pions in the fits produces a dramatic increase in the $\chi^2/dof$ 
value relative to the fits without the pions. Also the fitted temperature is 
calculated to be much higher in the presence of pions. This is checked from comparing
the value of $\chi^2/dof$ and the temperature in Tables (b) and (a). Although in the present work the fugacities
$\gamma_u$ and $\gamma_d$ are used, which describe off chemical equilobrium effects 
and the pion content is $u$ and $d$ quarks, this dramatic increase in the $\chi^2/dof$
and the temperature when the pions enter the fit, persists. Thus, for the QGP compatible cases, the
$\gamma_u$ and $\gamma_d$ fugacities cannot improve the fit to acceptable limits and
reveal that an excess of pions is present to these cases.

It is evident by incorporating all the previous analyses, that for the $S+S$, 
$S+Ag$ and $Au+Au$ cases, the freeze-out thermodynamic variables should be considered 
these of Table 3 (b). For the $Pb+Pb$ case the relevant freeze-out variables should be considered these of Table 2 (a) or 2 (b).

Gathering the observations made in this section, what it found is that the freeze-out
conditions of some experiments are compatible with a QGP phase (constraints B).
This is revealed from the fact that the imposition of constraints B produces fits with
acceptable values of $\chi^2/dof$ and temperature. The opposite is true for the
experiments incompatible with the QGP phase. In such cases the imposition of
constraints B leads to high values $\chi^2/dof$ and temperature, though these values are acceptable, when only constraints A (no QGP phase assumed) are imposed.
Also, the QGP compatible cases present an excess in the pion multiplicity which cannot 
properly be fitted (in these cases) by the $\gamma_u$ and $\gamma_d$ fugacities. This
is seen by a great increase in the value of $\chi^2/dof$ and temperature when the pions
are included in the fit in comparison to the fits without pions, for both cases of
constraints A and B. 

\vspace{0.3cm}
{\large \bf 6. Conclusions}

Although two different partition functions are used for the description
of the quark and hadronic side of matter, it is possible to preserve the
continuity of all chemical potentials and, of course, temperature and pressure (Gibbs equilibrium conditions) at the
transition region, which is confined to a curve. Also, all
the constraints imposed by the
conservation laws of quantum quantities can be
applied, leading, at the same time, to a non-trivial solution of the
thermodynamic variables into a three quark flavour system. The key issue
for the success of this project is the expansion of the fugacity sector
of the available variables and the, already, introduced relative chemical
equilibrium variables can be used to serve that purpose.

Despite the fact that the space of the thermodynamic variables is extended, the restrictions imposed on  the transition points produce relations among these variables. A part of these relations, in a simple form, is expressed
in eq.~(22) or (24).

The restrictions on the freeze-out conditions, imposed by the existence of
a quark-gluon state in the early stages after a collision experiment, can
be applied to every case where the thermalisation of the produced hadrons
has been proven. They can serve to separate the experiments compatible
with QGP state from those which are not.
The expansion of the fugacity sector with the partial equilibrium
fugacities, though, magnifies the part of the phase space allowed by such
constraints.

The whole methodology which was presented can be used for every grand
canonical partition function
adopted for the description of the HG or QGP phase. The inclusion of
interaction is crucial for the prediction of the critical point and the
volume expansion ratio, which could not be determined by the models used
in this work. At the moment, lattice calculations have led to the
determination
of the accurate quark-gluon equation of state with three quark flavours at
finite baryon chemical potential [20,31]. It would be interesting,
though, if these calculations could be extended with the inclusion of the
relative chemical equilibrium variables for light and strange quarks,
allowing for matching with the existing hadron gas models. For the hadronic side of matter the inclusion of the attractive part of interaction can be incorporated via the statistical bootstrap [7,8], where the prediction of a critical point
is also possible [32]. The incorporation of the full set of parameters
$\gamma_i$ to these studies would allow for a more precise matching with a
primordial quark phase.

{\bf Acknowledgement} I would like to thank N.~G.~Antoniou, C.~N.~Ktorides
and F.~K.~Diakonos for fruitful discussions.

\vspace{1cm}
{\large \bf Tables}
\vspace{1cm}

\begin{center}                                                                                                         
\begin{tabular}{|cc|cc|cc|cc|}
\hline
\multicolumn{2}{|c|}{$S+S$}           & \multicolumn{2}{|c|}{$S+Ag$}         & \multicolumn{2}{|c|}{$Pb+Pb$}         & \multicolumn{2}{|c|}{$Au+Au$}                 \\ \hline \hline
$K^+$                & $12.5\pm 0.4$  & ${K_s}^0$            & $15.5\pm 1.5$ & $N_p$                & $362\pm 5.1$   & $\Lambda$                  & $17.20\pm 1.75$  \\ 
$K^-$                & $6.9\pm 0.4$   & $\Lambda$            & $15.2\pm 1.2$ & $K^+$                & $103\pm 7.1$   & $\overline{\Lambda}$       & $12.15\pm 1.25$  \\
${K_s}^0$            & $10.5\pm 1.7$  & $\overline{\Lambda}$ & $2.6\pm 0.3$  & $K^-$                & $51.9\pm 3.6$  & $\Xi^-$                    & $2.11 \pm 0.23$  \\  
$\Lambda$            & $9.4\pm 1.0$   & $\overline{p}$       & $2.0\pm 0.8$  & ${K_s}^0$            & $81\pm 4$      & $\overline{\Xi}^+$         & $1.77 \pm 0.19$  \\ 
$\overline{\Lambda}$ & $2.2\pm 0.4$   & $p-\overline{p}$     & $43\pm 3$     & $\phi$               & $7.6\pm 1.1$   & $\Omega+\overline{\Omega}$ & $0.585\pm 0.150$ \\ 
$\overline{p}$       & $1.15\pm 0.40$ & $B-\overline{B}$     & $105\pm 12$   & $\Lambda$            & $53\pm 5$      & $p$                        & $26.37\pm 2.60$  \\ 
$p-\overline{p}$     & $21.2\pm 1.3$  & $h^{-(*)}$           & $186\pm 11$   & $\overline{\Lambda}$ & $4.64\pm 0.32$ & $\overline{p}$             & $18.72\pm 1.90$  \\
$B-\overline{B}$     & $54\pm 3$      &                      &               & $\Xi^-$              & $4.45\pm 0.22$ & ${K_s}^0$                  & $36.7 \pm 5.5$   \\ 
$h^{-(*)}$           & $98\pm 3$      &                      &               & $\overline{\Xi}^+$   & $0.83\pm 0.04$ & $\phi$                     & $5.73 \pm 0.78$  \\  
                     &                &                      &               & $\Omega$             & $0.62\pm 0.09$ & $K^{*0}$                   & $10.0 \pm 2.70$  \\  
                     &                &                      &               & $\overline{\Omega}$  & $0.20\pm 0.03$ & $\pi^{+(*)}$               & $239  \pm 10.6$  \\  
                     &                &                      &               & $\pi^{+(*)}$         & $619\pm 35.4$  & $\pi^{-(*)}$               & $239  \pm 10.6$  \\ 
                     &                &                      &               & $\pi^{-(*)}$         & $639\pm 35.4$  & $K^+/K^-$                  & $1.092\pm 0.023$ \\
                     &                &                      &               &                      &                & $\overline{K}^{*0}/K^{*0}$ & $0.92 \pm 0.27$  \\
                     &                &                      &               &                      &                & $\overline{\Omega}/\Omega$ & $0.95 \pm 0.16$  \\ \hline
\end{tabular}

{\footnotesize
$^{(*)}$ This multiplicity has not been used in the fits where the pions 
are excluded.} 
\end{center} 
 
\begin{center} 
Table 1. The full phase space multiplicities from the collision
experiments $S+S$ (NA35), $S+Ag$ (NA35) and $Pb+Pb$ (NA49), as well as
the midrapidity multiplicities and ratios from $Au+Au$ (STAR), used in the
fits. 
\end{center} 
\vspace{0.5cm}

\newpage
\begin{center}
(a) set A, Fit with all

\begin{tabular}{|c|c|c|c|c|}
\hline
               & $S+S$           & $S+Ag$          & $Pb+Pb$         & $Au+Au$           \\ \hline
\multicolumn{5}{|c|}{No pentaquarks}      
                  \\ \hline
$\chi^2/dof$   & $6.04/3$        & $4.18/1$        & $17.51/7$       & $13.62/9$         \\
$T\;(MeV)$     & $204.2\pm9.7$   & $236\pm19$      & $193.4\pm3.0$   & $326\pm76$        \\
$\lambda_u$    & $1.76\pm0.15$   & $1.77\pm0.33$   & $1.724\pm0.097$ & $1.20\pm0.12$     \\
$\lambda_d$    & $1.339\pm0.069$ & $1.47\pm0.19$   & $1.640\pm0.071$ & $1.039\pm0.014$   \\
$\lambda_s$    & $1.019\pm0.023$ & $1.020\pm0.041$ & $1.167\pm0.013$ & $1.0014\pm0.0047$ \\
$\gamma_u$     & $0.567\pm0.077$ & $0.46\pm0.14$   & $0.448\pm0.059$ & $0.164\pm0.091$   \\
$\gamma_d$     & $1.06\pm0.14$   & $0.70\pm0.17$   & $0.583\pm0.053$ & $0.606\pm0.084$   \\
$\gamma_s$     & $0.517\pm0.024$ & $0.338\pm0.043$ & $0.378\pm0.019$ & $0.284\pm0.069$   \\
$VT^3$         & $173\pm43$      & $350\pm150$     & $3671\pm400$    & $550\pm180$       \\
$\mu_B\;(MeV)$ & $234\pm29$      & $315\pm80$      & $297\pm21$      & $84\pm38	$        \\ \hline
\multicolumn{5}{|c|}{With pentaquarks}      
                  \\ \hline
$\chi^2/dof$   & $6.20/3$        & $4.11/1$        & $17.49/7$       & $13.62/9$         \\
$T\;(MeV)$     & $205\pm10$      & $235\pm19$      & $193.7\pm3.0$   & $326\pm76$        \\
$\lambda_u$    & $1.76\pm0.15$   & $1.77\pm0.33$   & $1.724\pm0.097$ & $1.20\pm0.12$     \\
$\lambda_d$    & $1.340\pm0.070$ & $1.46\pm0.19$   & $1.639\pm0.071$ & $1.039\pm0.014$   \\
$\lambda_s$    & $1.022\pm0.022$ & $1.022\pm0.040$ & $1.168\pm0.013$ & $1.0015\pm0.0047$ \\
$\gamma_u$     & $0.563\pm0.075$ & $0.46\pm0.14$   & $0.446\pm0.059$ & $0.164\pm0.091$   \\
$\gamma_d$     & $1.05\pm0.14$   & $0.70\pm0.16$   & $0.582\pm0.053$ & $0.606\pm0.084$   \\
$\gamma_s$     & $0.514\pm0.017$ & $0.340\pm0.042$ & $0.376\pm0.019$ & $0.284\pm0.069$   \\
$VT^3$         & $173\pm43$      & $350\pm150$     & $3670\pm400$    & $550\pm180$       \\
$\mu_B\;(MeV)$ & $235\pm30$      & $313\pm79$      & $297\pm21$      & $84\pm38$         \\ \hline
\end{tabular}
\end{center}

\newpage
\begin{center}
(b) set A, Fit with no $\pi$'s

\begin{tabular}{|c|c|c|c|c|}
\hline
               & $S+S$           & $S+Ag$          & $Pb+Pb$         & $Au+Au$           \\ \hline
\multicolumn{5}{|c|}{No pentaquarks}      
                  \\ \hline
$\chi^2/dof$   & $0.36/2$        & $0/0$           & $8.97/5$        & $1.83/8$          \\
$T(MeV)$       & $260\pm23$      & $259.3$         & $174.3\pm4.6$   & $151.6\pm3.4$     \\
$\lambda_u$    & $1.781\pm0.082$ & $1.6666$        & $1.697\pm0.060$ & $1.0756\pm0.0051$ \\
$\lambda_d$    & $1.514\pm0.066$ & $1.8180$        & $1.710\pm0.069$ & $1.0729\pm0.0055$ \\
$\lambda_s$    & $0.939\pm0.013$ & $0.9611$        & $1.169\pm0.012$ & $1.0172\pm0.0024$ \\
$\gamma_u$     & $0.548\pm0.027$ & $0.5819$        & $0.670\pm0.048$ & $1.37\pm0.14$     \\
$\gamma_d$     & $0.719\pm0.061$ & $0.5377$        & $0.759\pm0.055$ & $1.75\;{\rm(fixed)}$\\
$\gamma_s$     & $0.588\pm0.022$ & $0.4650$        & $0.575\pm0.063$ & $1.93\pm0.17$     \\
$VT^3$         & $67.3\pm9.2$    & $158.08$        & $2426\pm30$     & $275\pm23$        \\
$\mu_B(MeV)$   & $365\pm41$      & $442.4$         & $279\pm17$      & $32.4\pm1.9$      \\ \hline
\multicolumn{5}{|c|}{With pentaquarks}      
                  \\ \hline
$\chi^2/dof$   & $0.35/2$        & $0/0$           & $9.10/5$        & $1.85/8$          \\
$T(MeV)$       & $258\pm23$      & $265.1$         & $176.1\pm5.0$   & $151.8\pm3.4$     \\
$\lambda_u$    & $1.773\pm0.084$ & $1.6664$        & $1.700\pm0.062$ & $1.0761\pm0.0052$ \\
$\lambda_d$    & $1.522\pm0.080$ & $1.8210$        & $1.705\pm0.069$ & $1.0732\pm0.0055$ \\
$\lambda_s$    & $0.943\pm0.018$ & $0.9612$        & $1.170\pm0.012$ & $1.0178\pm0.0025$ \\
$\gamma_u$     & $0.557\pm0.071$ & $0.5676$        & $0.646\pm0.055$ & $1.36\pm0.13$     \\
$\gamma_d$     & $0.72\pm0.10$   & $0.5237$        & $0.741\pm0.054$ & $1.75\;{\rm(fixed)}$\\
$\gamma_s$     & $0.597\pm0.056$ & $0.4544$        & $0.555\pm0.065$ & $1.92\pm0.17$     \\
$VT^3$         & $67\pm14$       & $153.42$        & $2490\pm85$     & $274\pm23$        \\
$\mu_B(MeV)$   & $365\pm44$      & $453.1$         & $282\pm18$      & $32.6\pm1.9$      \\ \hline
\end{tabular}
\end{center}

\begin{center} 
Table 2. The results of fits on the $S+S$ (NA35), $S+Ag$ (NA35), $Pb+Pb$
(NA49) and $Au+Au$ (STAR) data with the imposition of the set of
constraints A, without and with the inclusion of the pentaquark states. 
In part (a) all the multiplicities have been included in the fit and in
part b) the multiplicities that contain pions have been excluded from the fit. In part (b) the fit on the STAR data has been performed with the
variable $\gamma_d$ held fixed at the given value.
Table 2 (a) or 2 (b) contains the extracted parameters which are concluded as the
freeze-out conditions for the $Pb+Pb$ case.
\end{center}

\newpage
\begin{center}
(a) set B, Fit with all

\begin{tabular}{|c|c|c|c|c|}
\hline
               & $S+S$           & $S+Ag$          & $Pb+Pb$         & $Au+Au$           \\ \hline
\multicolumn{5}{|c|}{No pentaquarks}      
                  \\ \hline
$\chi^2/dof$   & $14.76/5$       & $5.77/3$        & $162.13/9$      & $42.50/11$        \\
$T\;(MeV)$     & $243\pm26$      & $275\pm55$      & $437\pm32$      & $345\pm63$        \\
$\lambda_u$    & $1.536\pm0.038$ & $1.613\pm0.046$ & $1.668\pm0.084$ & $1.082\pm0.021$   \\
$\lambda_d$    & $1.534\pm0.038$ & $1.638\pm0.048$ & $1.728\pm0.095$ & $1.086\pm0.022$   \\
$\gamma_u$     & $0.58\pm0.12$   & $0.46\pm0.14$   & $0.273\pm0.019$ & $0.350\pm0.065$   \\
$\gamma_d$     & $0.58\pm0.12$   & $0.47\pm0.14$   & $0.293\pm0.021$ & $0.381\pm0.069$   \\
$\gamma_s$     & $0.401\pm0.094$ & $0.308\pm0.095$ & $0.188\pm0.011$ & $0.322\pm0.061$   \\
$VT^3$         & $152.2\pm9.5$   & $280\pm12$      & $678\pm77$      & $378\pm53$        \\
$\mu_B\;(MeV)$ & $313\pm36$      & $403\pm82$      & $702\pm74$      & $84\pm22$         \\
$V_0(10^{-11}MeV^{-4})$& $5.4\pm2.0$ & $4.9\pm1.4$ & $2.33\pm0.35$   & $3.58\pm0.62$     \\
$B^{1/4}\;(MeV)$&$325\pm37$      & $359\pm74$      & $548\pm40$      & $441\pm83$        \\ \hline
\multicolumn{5}{|c|}{With pentaquarks}      
                  \\ \hline
$\chi^2/dof$   & $14.99/5$       & $5.78/3$        & $162.65/9$      & $42.53/11$        \\
$T\;(MeV)$     & $246\pm22$      & $278\pm74$      & $439\pm32$      & $346\pm63$        \\
$\lambda_u$    & $1.536\pm0.038$ & $1.613\pm0.046$ & $1.668\pm0.084$ & $1.082\pm0.021$   \\
$\lambda_d$    & $1.535\pm0.038$ & $1.638\pm0.048$ & $1.728\pm0.095$ & $1.086\pm0.022$   \\
$\gamma_u$     & $0.58\pm0.10$   & $0.45\pm0.18$   & $0.272\pm0.019$ & $0.349\pm0.064$   \\
$\gamma_d$     & $0.58\pm0.10$   & $0.47\pm0.19$   & $0.293\pm0.021$ & $0.380\pm0.069$   \\
$\gamma_s$     & $0.397\pm0.079$ & $0.31\pm0.13$   & $0.188\pm0.011$ & $0.322\pm0.061$   \\
$VT^3$         & $149.1\pm8.5$   & $274\pm11$      & $672\pm76$      & $377\pm52$        \\
$\mu_B\;(MeV)$ & $317\pm32$      & $407\pm110$     & $705\pm74$      & $85\pm22$         \\
$V_0(10^{-11}MeV^{-4})$& $5.5\pm1.5$ & $4.9\pm1.6$ & $2.31\pm0.35$   & $3.57\pm0.63$     \\
$B^{1/4}\;(MeV)$& $328\pm31$     & $362\pm100$     & $551\pm41$      & $442\pm82$        \\ \hline
\end{tabular}
\end{center}

\newpage
\begin{center}
(b) set B, Fit with no $\pi$'s 

\begin{tabular}{|c|c|c|c|c|}
\hline
               & $S+S$           & $S+Ag$          & $Pb+Pb$         & $Au+Au$           \\ \hline
\multicolumn{5}{|c|}{No pentaquarks}      
                  \\ \hline
$\chi^2/dof$   & $1.88/4$        & $0.13/2$        & $127.72/7$      & $10.72/9$         \\
$T\;(MeV)$     & $194.9\pm4.4$   & $209.3\pm4.9$   & $444\pm38$      & $218\pm26$        \\
$\lambda_u$    & $1.605\pm0.025$ & $1.661\pm0.011$ & $1.746\pm0.073$ & $1.075\pm0.012$   \\
$\lambda_d$    & $1.599\pm0.025$ & $1.695\pm0.012$ & $1.817\pm0.083$ & $1.080\pm0.012$   \\
$\gamma_u$     & $0.949\pm0.065$ & $0.775\pm0.048$ & $0.261\pm0.018$ & $0.71\pm0.21$     \\
$\gamma_d$     & $0.958\pm0.067$ & $0.793\pm0.048$ & $0.280\pm0.020$ & $0.76\pm0.22$     \\
$\gamma_s$     & $0.847\pm0.071$ & $0.601\pm0.042$ & $0.199\pm0.012$ & $0.81\pm0.27$     \\
$VT^3$         & $94.1\pm6.5$    & $199\pm10$      & $621\pm67$      & $269\pm59$        \\
$\mu_B\;(MeV)$ & $275.3\pm9.3$   & $327.0\pm8.4$   & $778\pm79$      & $49.5\pm8.1$      \\
$V_0(10^{-11}MeV^{-4})$& $6.8\pm1.0$ & $6.2\pm0.8$ & $2.2\pm0.4$     & $5.4\pm4.6$       \\
$B^{1/4}\;(MeV)$& $278.9\pm7.3$  & $290.7\pm7.6$   & $557\pm48$      & $301\pm41$        \\ \hline
\multicolumn{5}{|c|}{With pentaquarks}      
                  \\ \hline
$\chi^2/dof$   & $1.55/4$        & $0.13/2$        & $128.04/7$      & $10.80/9$         \\
$T\;(MeV)$     & $196.7\pm4.6$   & $216.3\pm5.2$   & $446\pm38$      & $221\pm26$        \\
$\lambda_u$    & $1.608\pm0.023$ & $1.663\pm0.011$ & $1.746\pm0.073$ & $1.075\pm0.012$   \\
$\lambda_d$    & $1.603\pm0.023$ & $1.694\pm0.012$ & $1.817\pm0.083$ & $1.081\pm0.012$   \\
$\gamma_u$     & $0.948\pm0.069$ & $0.726\pm0.044$ & $0.261\pm0.018$ & $0.69\pm0.20$     \\
$\gamma_d$     & $0.956\pm0.070$ & $0.746\pm0.045$ & $0.280\pm0.020$ & $0.74\pm0.21$     \\
$\gamma_s$     & $0.865\pm0.078$ & $0.562\pm0.039$ & $0.198\pm0.012$ & $0.79\pm0.25$     \\
$VT^3$         & $88.5\pm6.9$    & $200\pm10$      & $615\pm66$      & $272\pm58$        \\
$\mu_B\;(MeV)$ & $279.1\pm9.1$   & $338.2\pm8.7$   & $781\pm80$      & $50.4\pm8.2$      \\
$V_0(10^{-11}MeV^{-4})$& $7.2\pm0.9$ & $6.3\pm0.7$ & $2.2\pm0.4$     & $5.4\pm4.3$       \\
$B^{1/4}\;(MeV)$& $282.1\pm7.5$  & $298.4\pm7.8$   & $560\pm48$      & $305\pm40$        \\ \hline
\end{tabular}
\end{center}

\begin{center} 
Table 3. The results of fits on the $S+S$ (NA35), $S+Ag$ (NA35), $Pb+Pb$
(NA49) and $Au+Au$ (STAR) data with the imposition of the set of
constraints B, without and with the inclusion of the pentaquark states. 
In part (a) all the multiplicities have been included in the fit and in
part (b) the multiplicities that contain pions have been excluded from the fit.
Table 3 (b) contains the extracted parameters which are concluded as the
freeze-out conditions for the $S+S$, $S+Ag$ and $Au+Au$ cases.
\end{center} 

\newpage
{\bf Figure Captions}
\newtheorem{f}{Fig.} 
\begin{f} 
\rm Temperature as a function of the baryon chemical potential at the
QGP-Hadron gas transition line for $B^{1/4}=280$ MeV, without (lines
(a),(c)) and with (line (b)) the inclusion of the pentaquark states. Lines (a) and (b) are calculated with $V_0=1.4/(4B)$ and line (c) with $V_0=1.29/(4B)$. Line (d) is the phase transition curve calculated from lattice QCD in [21].
\end{f}
\begin{f} 
\rm Relative chemical equilibrium variable of $u$-quark, $\gamma_u$, as
a function of the baryon chemical potential at the QGP-Hadron gas
transition line. Lines (a)-(c) correspond to lines (a)-(c) of Fig.~1. 
\end{f}
\begin{f} 
\rm Relative chemical equilibrium variable of $d$-quark, $\gamma_d$, as
a function of relative chemical equilibrium variable of $u$-quark,
$\gamma_u$, at the QGP-Hadron gas transition line for the isospin
symmetric case. Lines (a)-(c) correspond to lines (a)-(c) of Fig.~1. The
line $\gamma_d=\gamma_u$ is also depicted.
\end{f}
\begin{f}
\rm Relative chemical equilibrium variable of $s$-quark, $\gamma_s$, as
a function of the baryon chemical potential at the QGP-Hadron gas
transition line. Lines (a)-(c) correspond to lines (a)-(c) of Fig.~1.
\end{f}
\begin{f}
\rm Volume expansion ratio $x=V_{HG}/V_{QGP}$ between pure hadron and pure QGP phase at the same transition point, as a function of the baryon
chemical potential, which was used in calculations in Figs.~1-4. Lines
(a)-(c) correspond to lines (a)-(c) of Fig.~1. In lines (a) and (b)
$x_1=1.1$ at $\lambda_{u\;1}=14.2$ and $\epsilon=1.5$ and in line (c) $x_1=1.12$ at $\lambda_{u\;1}=14.2$ and $\epsilon=1.15$.
\end{f}

\newpage
\pagestyle{empty}

\vspace{-0.5cm}\hspace{-0.7cm}\includegraphics[scale=0.95,angle=90]{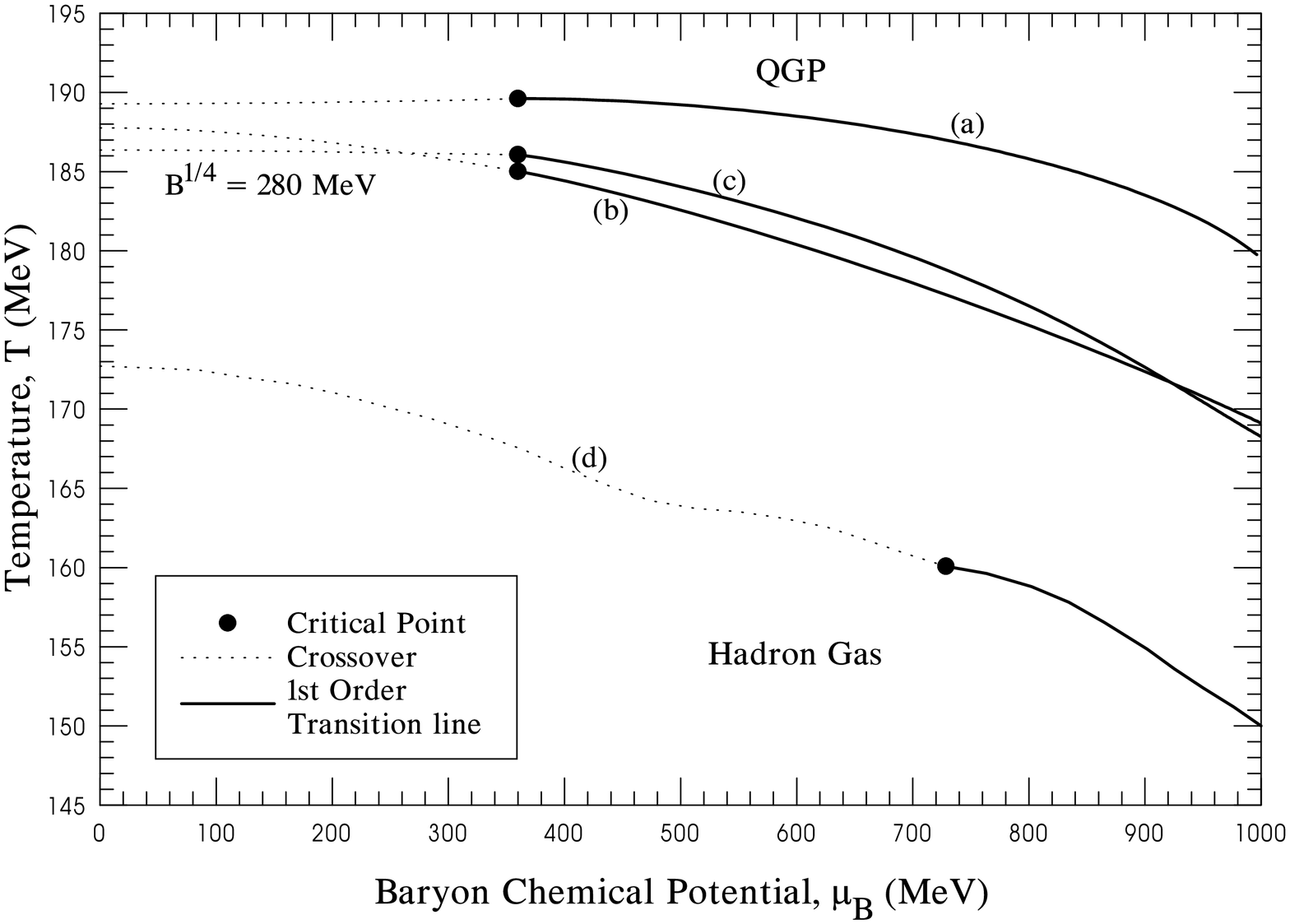}
\begin{center}
Fig. 1
\end{center}

\vspace{-0.5cm}\hspace{-0.7cm}\includegraphics[scale=0.95,angle=90]{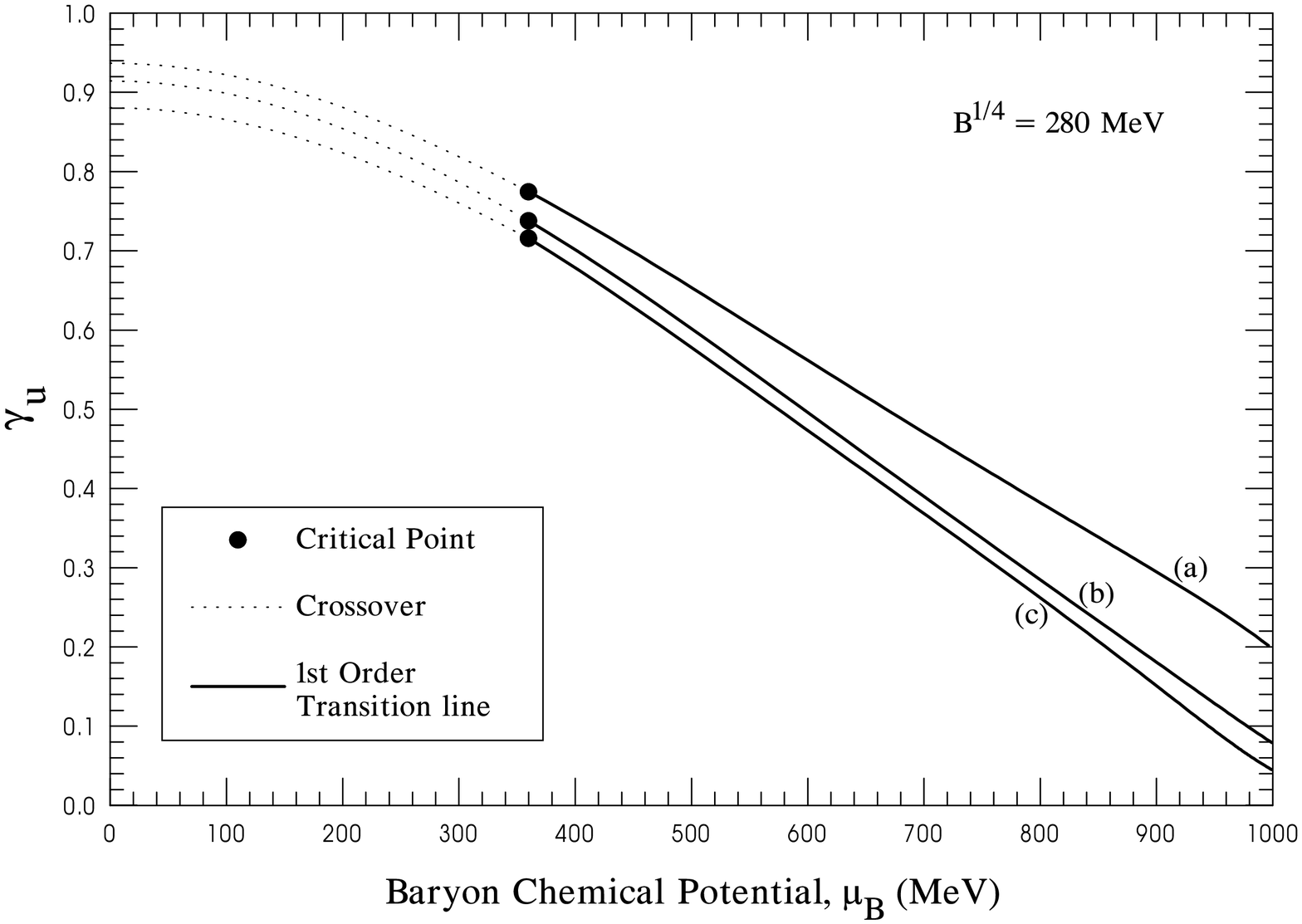}
\begin{center}
Fig. 2
\end{center}

\vspace{-0.5cm}\hspace{-0.7cm}\includegraphics[scale=0.95,angle=90]{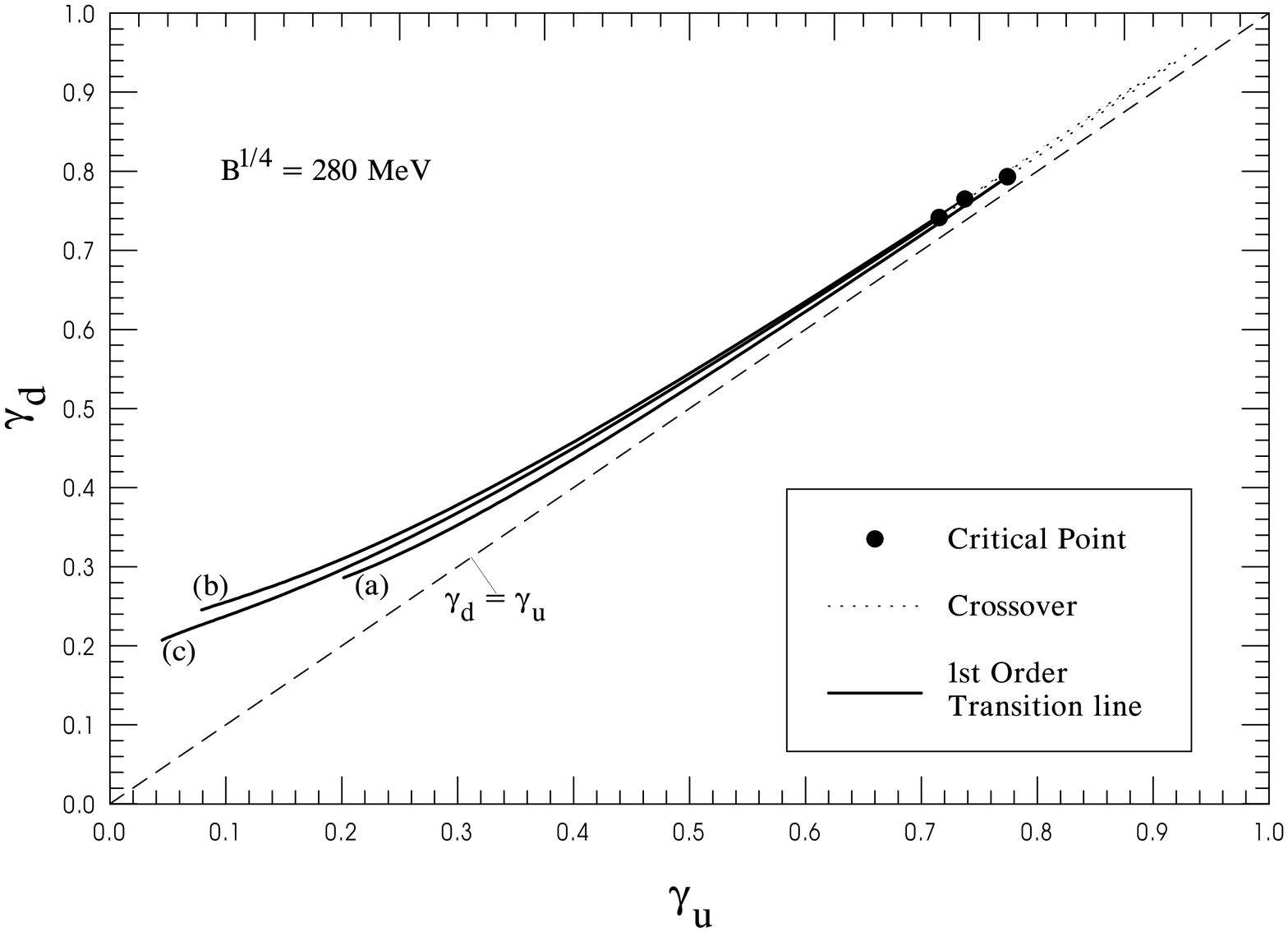}
\begin{center}
Fig. 3
\end{center}

\vspace{-0.5cm}\hspace{-0.7cm}\includegraphics[scale=0.95,angle=90]{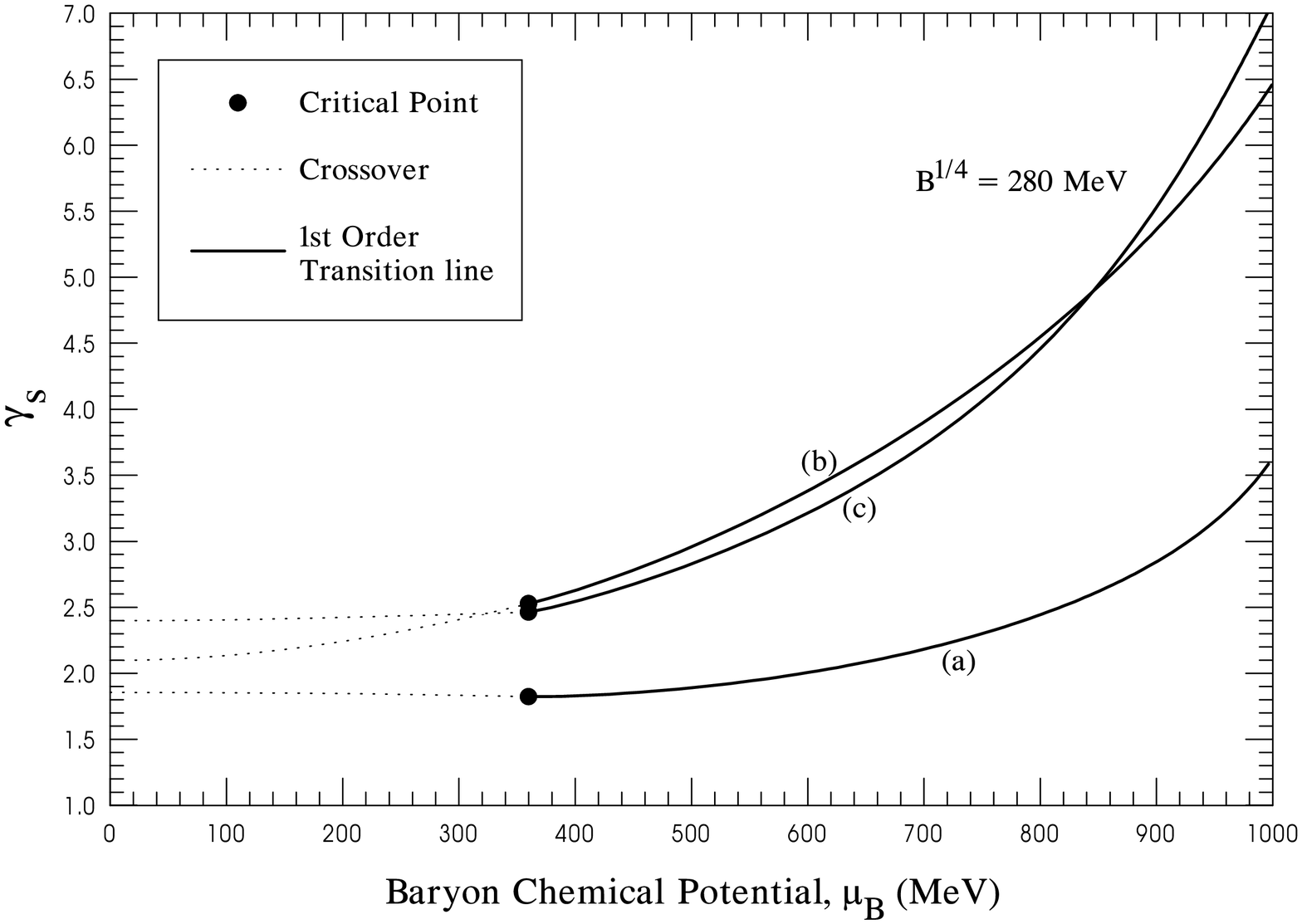}
\begin{center}
Fig. 4
\end{center}

\vspace{-0.5cm}\hspace{-0.7cm}\includegraphics[scale=0.95,angle=90]{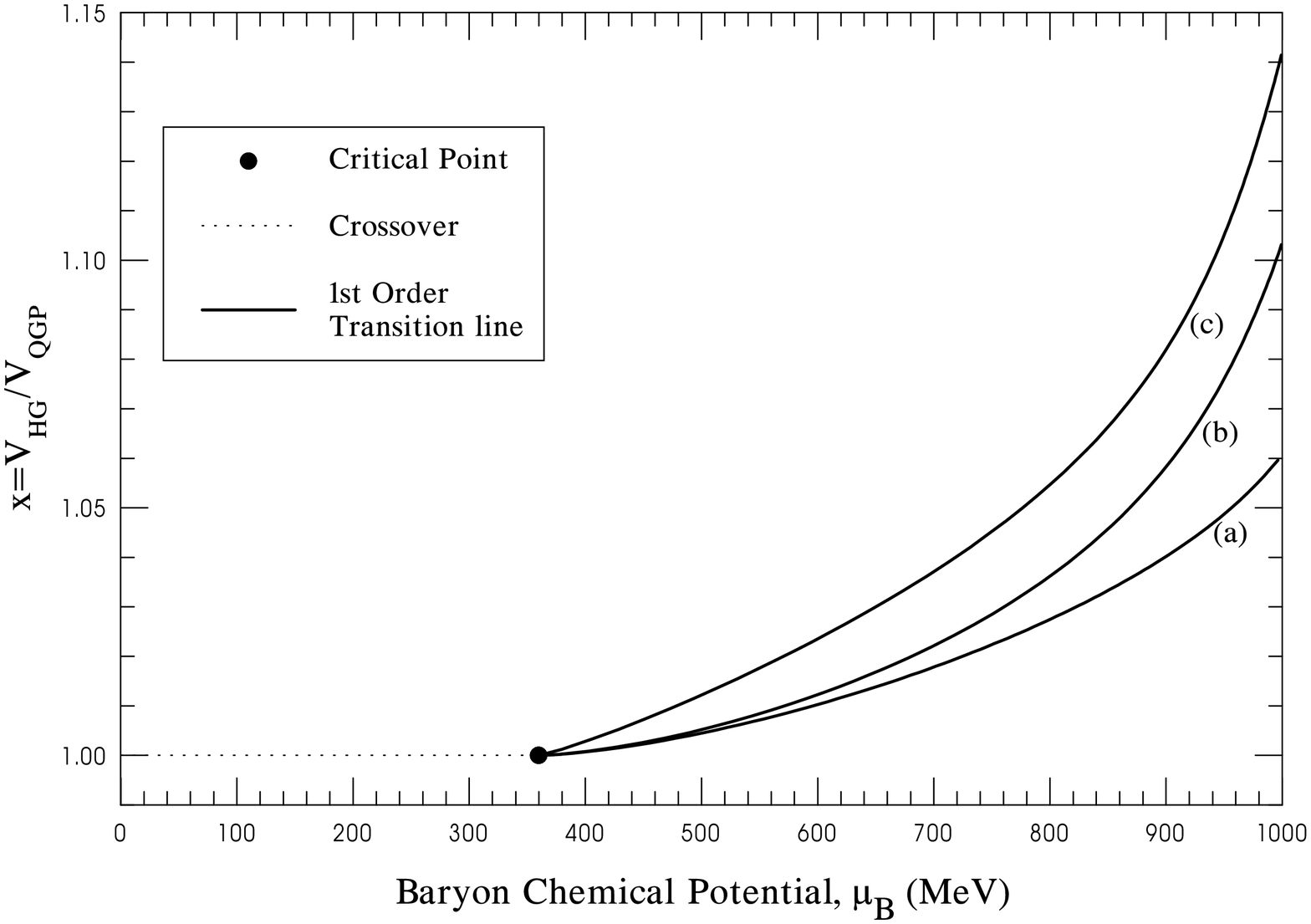}
\begin{center}
Fig. 5
\end{center}

\end{document}